# A Molecular Dynamics Study of Mechanical Properties of Vertically Stacked Silicene/MoS$_2$ van der Waals Heterostructure

Bishawjit Kar[1], Plabon Paul[1], Md Arshadur Rahman[1], Mohammad Jane Alam Khan[1,a)]

[1]Department of Mechanical Engineering, Bangladesh University of Engineering and Technology, Dhaka-1000, Bangladesh

a) Corresponding author: ronin@me.buet.ac.bd

**Abstract.** Silicene is an intriguing silicon allotrope with a honeycomb lattice structure similar to graphene with slightly buckled geometry. Molybdenum disulfide (MoS$_2$), on the other hand, is a significant 2D transition metal dichalcogenide that has demonstrated promise in a variety of applications. Van der Waals heterostructures, which are created by stacking distinct 2D crystals on top of each other, are becoming increasingly important due to their unique optoelectronic and electromechanical properties. Using molecular dynamics simulations, the mechanical characteristics of vertically stacked Silicene/MoS$_2$ van der Waals heterostructures are examined in this study. The response and structural stability of the heterostructures at various loading orientations and temperatures are given particular attention. The research findings highlight that the fracture strength of the Silicene/MoS$_2$ heterostructure decreases by 40% in both armchair and zigzag orientations when the temperature is raised from 100K to 600K. Furthermore, a linear decrease in Young's modulus is observed as temperature rises. It is noteworthy that the Rule of Mixture (ROM) predictions for Young's Moduli are observed to be marginally lower than the simulation results. The analyses reveal that the silicene layer fractures first under both loading directions shows crack propagation at ±60° in the armchair and predominantly perpendicular in zigzag, followed by subsequent MoS$_2$ layer failure. The study also shows that the MoS$_2$ layer largely determines the elastic properties of the heterostructure, whereas the silicene layer primarily dictates the failure of the heterostructure. These findings offer an in-depth understanding of the mechanical properties of Silicene/MoS$_2$ heterostructures, with significant implications for their use in cutting-edge nanoelectronics and nanomechanical systems.

## 1. INTRODUCTION

The beginning of the 21st century has brought about a significant shift in the field of materials science, characterised by the appearance of two-dimensional (2D) materials. These materials have had a profound impact on the discipline of materials engineering, fundamentally altering its landscape. The discovery of graphene in 2004 was a significant breakthrough in the field of materials science. Graphene is a representative example of a two-dimensional substance, consisting of a single layer of carbon atoms that are organized in a hexagonal lattice structure [1]. The remarkable characteristics of this material sparked widespread enthusiasm across scientific and industrial communities, elevating its status as a versatile substance with significant potential. The unique characteristics of graphene, including its amazing electrical conductivity, mechanical strength, and thermal properties, have led to a wide range of potential applications in several fields, including electronics, optics, sensors, energy storage, and biodevices [2], [3]. However, the ongoing search for materials exhibiting even more exceptional features continues, driven by the inherent constraints of graphene, particularly its absence of a bandgap. This limitation hindered its use in specific electrical and optoelectronic devices.

The pursuit of highly advanced two-dimensional (2D) materials has prompted extensive investigation into a wide range of alternatives to graphene. These alternatives encompass hexagonal boron nitride, graphitic carbon nitride, silicene, germanene, and stanene, as well as transition metal dichalcogenides (TMDs) like molybdenum disulfide (MoS$_2$) [4]–[6]. These materials, characterized by their unique characteristics and structural qualities, have emerged as viable candidates in the continuously advancing field of two-dimensional (2D) materials.

Silicene is an exceptional silicon allotrope among the various candidates, with a honeycomb lattice structure similar to that of graphene. Nevertheless, it sets itself apart by exhibiting a subtle buckling in its atomic structure, which arises from the displacement of silicon atoms in the out-of-plane direction. Silicene emerges as a highly promising two-dimensional semiconductor with the ability to customize its band gap [7]. Simultaneously, molybdenum disulfide ($MoS_2$), which falls under the category of transition metal dichalcogenides, has emerged as a significant participant in the field of two-dimensional materials. $MoS_2$ exhibits a departure from the semimetallic properties of graphene due to its possession of a non-zero bandgap. Hence, it strategically situates itself in a favorable position for utilization in transistor technology. The inherent semiconducting properties, adaptable band gap tunability, and resilient mechanical characteristics of $MoS_2$ render it a compelling option for high-performance electronic and nano-mechanical systems. It can be effectively employed in the development of low-power field-effect transistors that exhibit a high current switch ratio, elevated electron mobility, and substantial on-current density [8], [9].

Van der Waals heterostructures offer exciting opportunities in 2D materials research by enabling custom engineering through vertical stacking, leveraging weak van der Waals interactions. With favorable lattice mismatch ratios, these structures expand 2D material applications and form a foundational platform for future electronic and energy conversion devices. Notably, while electrical and optical properties have been extensively studied, the mechanical behavior of 2D lateral heterostructures remains underexplored. This study examines the mechanical characteristics of the Silicene/$MoS_2$ van der Waals heterostructure. The purpose of this study is to illuminate this structure's structural integrity and stability by thoroughly comprehending its mechanical characteristics. The lessons learned from this work will be used to highlight the possible uses of this heterostructure within the dynamic world of 2D materials.

## 2. COMPUTATIONAL METHOD

In order to construct the Silicene/$MoS_2$ heterostructure, individual structures for silicene and $MoS_2$ were initially generated using VESTA software [10]. The size of the silicene and $MoS_2$ unit cells are 3.83 Å and 3.19Å, respectively. The hexagonal systems were transformed into rectangular cells, yielding unit cell dimensions of 6.687 Å and 10.804 Å for silicene in the armchair and zigzag directions, and 3.861 Å and 12.475 Å for $MoS_2$ in the respective directions. To account for the lattice mismatch between silicene and $MoS_2$, a heterostructure was created by arranging 8x10 unit cells of $MoS_2$ and 16x10 unit cells of silicene and stacking them on top of each other (Fig 1). The silicene layer was slightly compressed, while $MoS_2$ experienced slight tension.

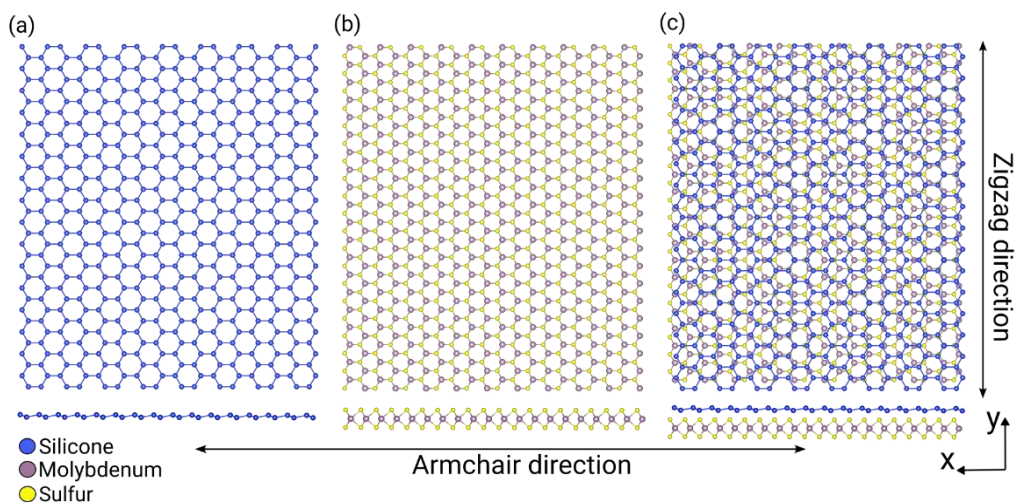

**FIGURE 1:** Geometry of (a) Single layer Silicene, (b) Single layer $MoS_2$, (c) Silicene- $MoS_2$ heterostructure.

The Molecular Dynamics simulations were conducted using the LAMMPS (Large-scale Atomic/Molecular Massively Parallel Simulator) program [11]. The interatomic interactions within Silicene and $MoS_2$ layers were effectively represented through the utilization of Many-body Stillinger-Weber potentials, parameterized by Jhiang et al. [12]. The first principles and experimental results were compared with the MD-computed effective Young's modulus for single-layer silicene and $MoS_2$, as shown in Table 1, to validate the interatomic potential. From Table 1, the effective Young's modulus for 2D materials can be defined as $YM_{effective} = YM \times thickness$. The thicknesses of silicene, $MoS_2$, and Silicene/$MoS_2$ are considered as 0.45 nm, 0.65 nm and 1.1 nm. These values are estimated assuming the atomic van der Waals radius and buckling height of silicene.

TABLE 1: Comparison of Effective Young's Modulus of silicene and $MoS_2$ with existing literature.

|  | This Study | | Literature | |
|---|---|---|---|---|
|  | Armchair (N/m) | Zigzag (N/m) | Armchair (N/m) | Zigzag (N/m) |
| **Silicene** | 62.53 | 62.20 | 63.50 [13] | 60.0 [13] |
| **$MoS_2$** | 108.26 | 108.36 | 120 ± 30 [14] | 120 ± 30 [14] |

The 12-6 Lennard-Jones (LJ) potential was considered for modeling the interlayer van der Waals interaction [15]. The two-body LJ interaction parameters were taken from the Universal Force Field (UFF) using the Lorentz-Berthelot mixing rule with the cut-off distance selected to be 14.0Å [16]. The LJ potential parameters used are, $\epsilon_{Si-Mo} = 6.506\ eV$, $\sigma_{Si-Mo} = 6.506$ Å and $\epsilon_{Si-S} = 6.506\ eV$, $\sigma_{Si-S} = 6.506$ Å and the potential energy can be expressed as,

$$E(r) = 4\epsilon \left[ \left(\frac{\sigma}{r}\right)^{12} - \left(\frac{\sigma}{r}\right)^{6} \right] \quad (1)$$

All simulation timesteps were set to 1 fs. The system was at first energy minimized using the conjugate gradient scheme. Then temperature equilibration was performed for 30 ps using a Berendsen thermostat with NVE ensemble. After that pressure and temperature equilibration were performed for another 30 ps under an NPT ensemble utilizing Nose/Hoover thermostat and Nose/Hoover barostat. Uniaxial tensile load was then applied in the armchair and zigzag directions under NPT at specified temperature, with the stress being measured using the virial stress theorem [17]. A strain rate of $10^9$ s$^{-1}$ was used for tensile loading. All simulation results were visualized using OVITO software [18].

## 3. RESULTS AND DISCUSSION

### 3.1. Temperature sensitivity and the influence of chirality on mechanical properties:

Temperature and chirality strongly influence the mechanical properties of a Silicene/$MoS_2$ heterostructure. Fig 2(a) and Fig 2(b) display stress-strain variations under uniaxial tensile load for temperatures of 100 K to 600 K at armchair and zigzag loading direction. Initially, the structure exhibits elastic deformation following Hooke's Law, with a linear relationship between stress and strain. As strain increases, stress becomes nonlinear, reaching the ultimate tensile strength (UTS) or fracture strength of the silicene layer. Another distinct peak in the stress-strain curve indicates the failure of the $MoS_2$ layer, revealing a brittle-type fracture without a ductile-to-brittle transition temperature. Examining the stress strain curves of both figures reveal that the Mos2 layer experiences greater strain following the initial failure of the silicene layer in armchair loading compared to zigzag loading. In order to examine chirality and temperature effects on Silicene/$MoS_2$ heterostructure, simulations were conducted for armchair and zigzag loading. In Fig 2(c) and Fig 2(d), results show that as temperature increases, fracture strength and strain decrease in both loading directions. The zigzag loading exhibits higher fracture strength but lower strain at failure compared to armchair. Fracture toughness, determined by the stress-strain curve area, consistently decreases with rising temperature. This decline is attributed to weakened atomic bonds due to increased lattice vibration and bond elongation at higher temperatures. From tensile stress analysis at 300K, it is observed that Silicene/$MoS_2$ has lower fracture strength and strain compared to both monolayers [Fig 2(e) & Fig 2(f)].

### 3.2. Effects on Young's Modulus:

Fig 3(a) reveals that Young's modulus (YM) decreases with rising temperature in the heterostructure, with the armchair direction exhibiting slightly higher YM values than the zigzag direction. Fig 3(b) and Fig 3(c) exhibit similar trends relating to Young's modulus (YM) and temperature for silicene, $MoS_2$ monolayers, and the Silicene/$MoS_2$ heterostructure. Notably, the YM of the heterostructure closely resembles that of $MoS_2$. To understand individual monolayer contributions to the heterostructure's elasticity, we used the Rule of Mixture (ROM) formula to estimate Young's modulus. The formula for the ROM used, $YM_{Si/MoS_2} = YM_{Si}f_{si} + YM_{MoS_2}f_{MoS_2}$, where $f_{si}$ and $f_{MoS_2}$ are volume fractions of silicene and $MoS_2$ monolayers respectively. The volume fraction for $MoS_2$, $f_{MoS_2} = 0.59$ is larger than the volume fraction of the silicene layer, $f_{Si} = 0.41$. Therefore, it is reasonable that $MoS_2$ would contribute more to the overall Young's modulus of the heterostructure.

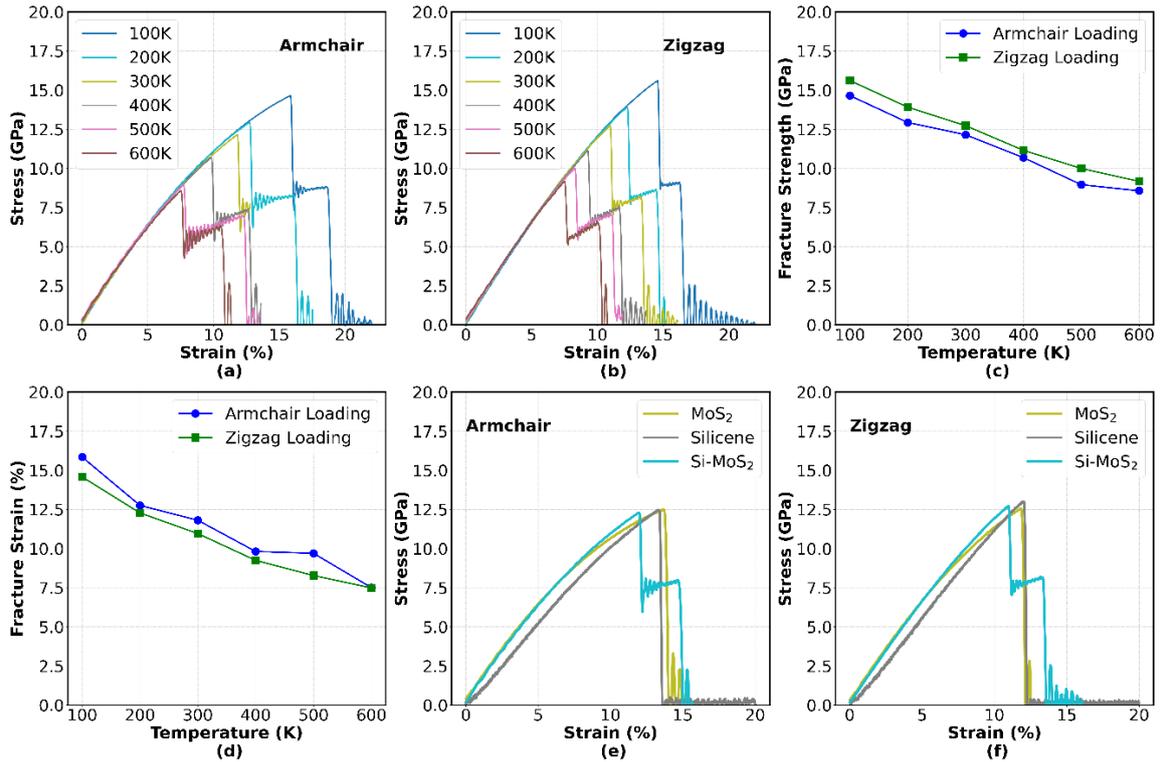

**FIGURE 2:** Variation of tensile stress with strain at different temperatures along (a) Armchair and (b) Zigzag loading direction are shown. Variation of (c) fracture strength and (d) fracture strain with temperature is also shown. Variation of tensile stress with strain for Silicene, $MoS_2$ monolayers and the Silicene/$MoS_2$ heterostructure with temperature are shown along (e) Armchair and (f) Zigzag loading direction.

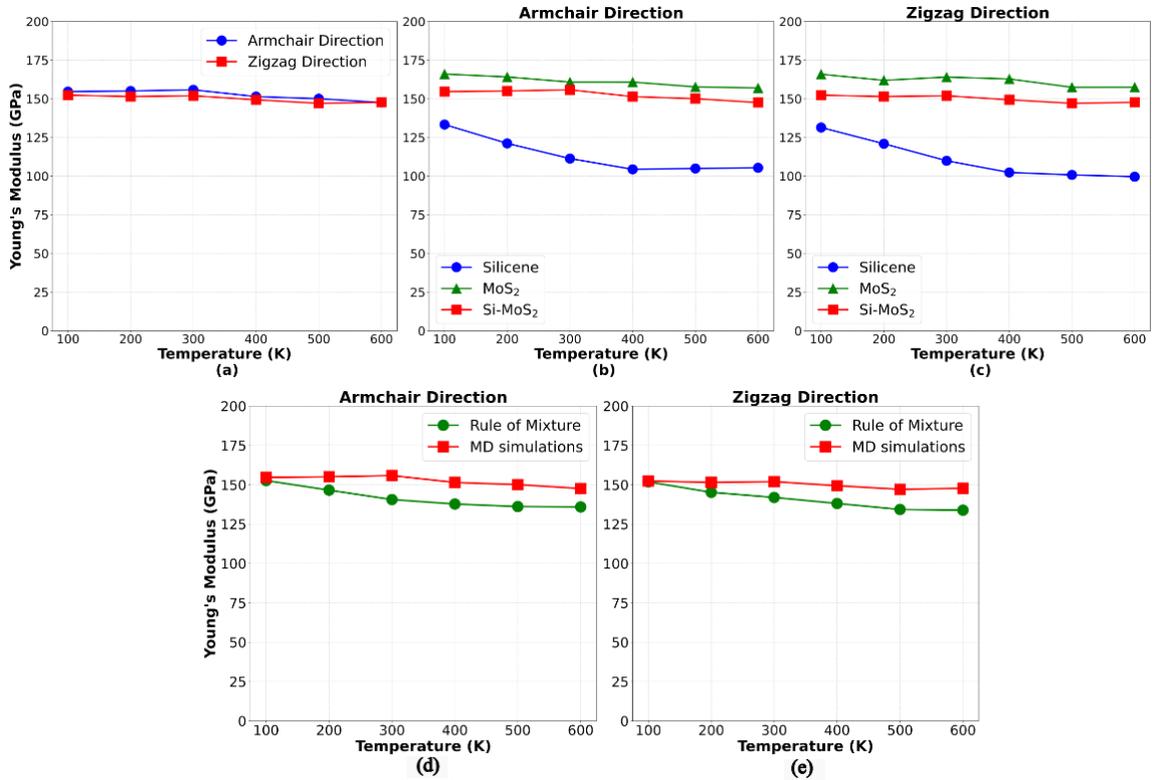

**FIGURE 3:** Variation of Young's Modulus with temperature for (a) Armchair and Zigzag Direction, (b) Silicene, $MoS_2$, and Silicene/$MoS_2$ along Armchair direction, (c) Silicene, $MoS_2$, and Silicene/$MoS_2$ along Zigzag direction, (d) MD results and ROM results along the Armchair direction. (e) MD results and ROM results along the Zigzag direction.

From Fig 3(d) and Fig 3(e), it is observable that, at lower temperatures, the outcomes obtained from both simulation and ROM calculations display a high degree of similarity. A divergence is detected in both directions, though, as the temperature rises. The largest deviation occurs at a temperature of 600K, with a value of around 11% for the armchair configuration and 9% for the zigzag configuration.

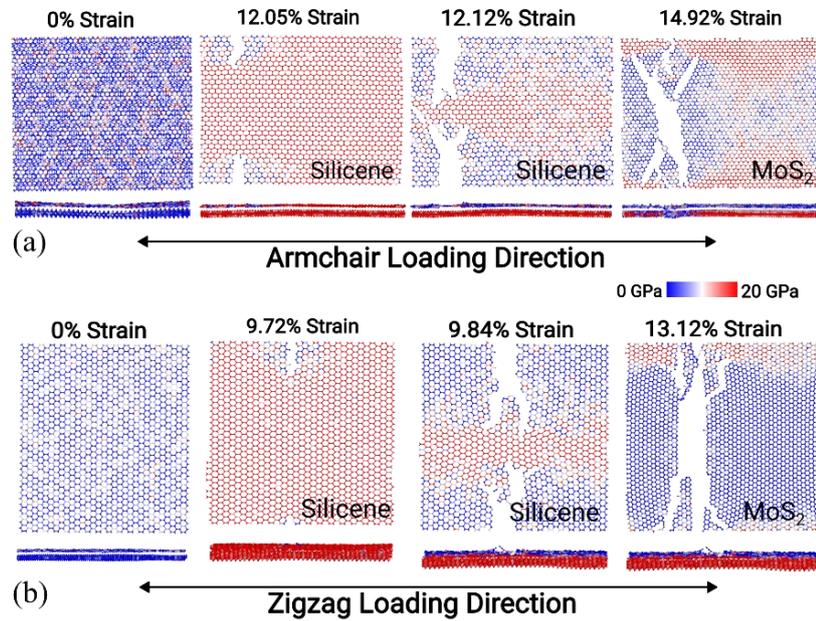

**FIGURE 4:** Fracture process of Silicene/$MoS_2$ heterostructure along (a) Armchair direction, (b) Zigzag direction at different strains for 300K temperature. In all cases, fracture originates from the Silicene layer.

### 3.3. Fracture Analysis:

Fig 4 provides insights into the fracture behavior of the heterostructure when subjected to armchair and zigzag loading at 300 K. In both loading scenarios, silicene exhibits a lower fracture strain than $MoS_2$, with the initial failure occurring in the silicene layer, followed by the eventual failure of the $MoS_2$ layer. The first appearance of cracks in the silicene layer under armchair loading [Fig 4(a)] occurs at a strain of roughly 12%. The cracks branch at an angle of ±60° relative to the loading direction, eventually leading to the failure of the silicene layer. The load then transfers to the $MoS_2$ layer, which cracks at around 15% strain with an analogous crack branching behavior. However, when subjected to zigzag loading [Fig 4(b)], the crack initiates in the silicene layer at 10% strain and after that the load shifts to the $MoS_2$ layer, which ultimately experiences fracture at around 13% strain. In this case, the crack propagates mostly perpendicularly to the loading direction.

The difference in crack behavior can be attributed to the fact that in armchair loading, there are two neighboring bonds near the crack tip at ±60° relative to the loading direction, which facilitates branching at these specific angles. However, in zigzag loading, the crack interacts with bonds oriented at right angles to the loading direction, impeding branching. Although occasional branching may occur due to thermal vibration. A similar phenomenon was observed in previous investigations related to other honeycomb-structured materials [19], [20].

## 4. CONCLUSION

This study illuminates Silicene/$MoS_2$ heterostructure behavior, enabling their use in next-generation nanoscale devices and systems. Different mechanical and fracture properties of the heterostructure have been examined using molecular dynamics along both armchair and zigzag loading orientations. The computational results indicate that the fracture strength and Young's modulus of the heterostructure decrease with increasing temperature. The zigzag loading direction exhibits higher fracture strength and Young's modulus but lower fracture strain compared to the armchair direction. It can be concluded that the silicene layer controls the failure behavior of the structure, whereas the elastic properties are determined primarily by the $MoS_2$ layer. Under armchair and zigzag loading, the Silicene/$MoS_2$ heterostructure shows initial failure in silicene with distinct crack patterns, followed by $MoS_2$ layer failure, influenced by bond orientations.

# 5. REFERENCES

[1] K. S. Novoselov et al., "Electric Field Effect in Atomically Thin Carbon Films," *Science*, vol. 306, no. 5696, pp. 666–669, Oct. 2004, doi: 10.1126/science.1102896.

[2] N. Savage, "Materials science: Super carbon," *Nature*, vol. 483, no. 7389, Art. no. 7389, Mar. 2012, doi: 10.1038/483S30a.

[3] N. Wei, L. Xu, H.-Q. Wang, and J.-C. Zheng, "Strain engineering of thermal conductivity in graphene sheets and nanoribbons: a demonstration of magic flexibility," *Nanotechnology*, vol. 22, no. 10, p. 105705, Mar. 2011, doi: 10.1088/0957-4484/22/10/105705.

[4] A. García-Miranda Ferrari, S. J. Rowley-Neale, and C. E. Banks, "Recent advances in 2D hexagonal boron nitride (2D-hBN) applied as the basis of electrochemical sensing platforms," *Anal. Bioanal. Chem.*, vol. 413, no. 3, pp. 663–672, Jan. 2021, doi: 10.1007/s00216-020-03068-8.

[5] M. Ezawa et al., "Fundamentals and functionalities of silicene, germanene, and stanene," *Riv. Nuovo Cimento*, vol. 41, no. 3, pp. 175–224, Mar. 2018, doi: 10.1393/ncr/i2018-10145-y.

[6] P. Malakar, M. S. H. Thakur, S. M. Nahid, and M. M. Islam, "Data-Driven Machine Learning to Predict Mechanical Properties of Monolayer Transition-Metal Dichalcogenides for Applications in Flexible Electronics," *ACS Appl. Nano Mater.*, vol. 5, no. 11, pp. 16489–16499, Nov. 2022, doi: 10.1021/acsanm.2c03564.

[7] "Tunable Bandgap in Silicene and Germanene | Nano Letters." Accessed: Sep. 27, 2023. [Online]. Available: https://pubs.acs.org/doi/10.1021/nl203065e

[8] S. Ahmed and J. Yi, "Two-Dimensional Transition Metal Dichalcogenides and Their Charge Carrier Mobilities in Field-Effect Transistors," *Nano-Micro Lett.*, vol. 9, no. 4, p. 50, Aug. 2017, doi: 10.1007/s40820-017-0152-6.

[9] T. Cui et al., "Mechanical reliability of monolayer $MoS_2$ and $WSe_2$," *Matter*, vol. 5, no. 9, pp. 2975–2989, Sep. 2022, doi: 10.1016/j.matt.2022.06.014.

[10] K. Momma and F. Izumi, "VESTA 3 for three-dimensional visualization of crystal, volumetric and morphology data," *J. Appl. Crystallogr.*, vol. 44, no. 6, Art. no. 6, Dec. 2011, doi: 10.1107/S0021889811038970.

[11] A. P. Thompson et al., "LAMMPS - a flexible simulation tool for particle-based materials modeling at the atomic, meso, and continuum scales," *Comput. Phys. Commun.*, vol. 271, p. 108171, Feb. 2022, doi: 10.1016/j.cpc.2021.108171.

[12] J.-W. Jiang and Y.-P. Zhou, "Parameterization of Stillinger-Weber Potential for Two- Dimensional Atomic Crystals," in *Handbook of Stillinger-Weber Potential Parameters for Two-Dimensional Atomic Crystals*, J.-W. Jiang and Y.-P. Zhou, Eds., InTech, 2017. doi: 10.5772/intechopen.71929.

[13] B. Mortazavi, O. Rahaman, M. Makaremi, A. Dianat, G. Cuniberti, and T. Rabczuk, "First-principles investigation of mechanical properties of silicene, germanene and stanene," *Phys. E Low-Dimens. Syst. Nanostructures*, vol. 87, pp. 228–232, Mar. 2017, doi: 10.1016/j.physe.2016.10.047.

[14] R. C. Cooper, C. Lee, C. A. Marianetti, X. Wei, J. Hone, and J. W. Kysar, "Nonlinear elastic behavior of two-dimensional molybdenum disulfide," *Phys. Rev. B*, vol. 87, no. 3, p. 035423, Jan. 2013, doi: 10.1103/PhysRevB.87.035423.

[15] J. E. Lennard-Jones and A. F. Devonshire, "Critical phenomena in gases - I," *Proc. R. Soc. Lond. Ser. - Math. Phys. Sci.*, vol. 163, no. 912, pp. 53–70, Jan. 1997, doi: 10.1098/rspa.1937.0210.

[16] A. K. Rappe, C. J. Casewit, K. S. Colwell, W. A. Goddard, and W. M. Skiff, "UFF, a full periodic table force field for molecular mechanics and molecular dynamics simulations," *J. Am. Chem. Soc.*, vol. 114, no. 25, pp. 10024–10035, Dec. 1992, doi: 10.1021/ja00051a040.

[17] J. A. Zimmerman, E. B. WebbIII, J. J. Hoyt, R. E. Jones, P. A. Klein, and D. J. Bammann, "Calculation of stress in atomistic simulation," *Model. Simul. Mater. Sci. Eng.*, vol. 12, no. 4, p. S319, Jun. 2004, doi: 10.1088/0965-0393/12/4/S03.

[18] A. Stukowski, "Visualization and analysis of atomistic simulation data with OVITO–the Open Visualization Tool," *Model. Simul. Mater. Sci. Eng.*, vol. 18, no. 1, p. 015012, Dec. 2009, doi: 10.1088/0965-0393/18/1/015012.

[19] T. M. T. Oishi, P. Malakar, M. Islam, and M. M. Islam, "Atomic-scale perspective of mechanical properties and fracture mechanisms of graphene/WS2/graphene heterostructure," *Comput. Condens. Matter*, vol. 29, p. e00612, Dec. 2021, doi: 10.1016/j.cocom.2021.e00612.

[20] E. H. Chowdhury, Md. H. Rahman, S. Fatema, and M. M. Islam, "Investigation of the mechanical properties and fracture mechanisms of graphene/WSe2 vertical heterostructure: A molecular dynamics study," *Comput. Mater. Sci.*, vol. 188, p. 110231, Feb. 2021, doi: 10.1016/j.commatsci.2020.110231.